\documentclass[review]{elsarticle}

\usepackage{hyperref}
\usepackage{subcaption}
\journal{Astronomy and Computing}
\usepackage{graphicx}

\usepackage{listings}

\begin{document}
\begin{frontmatter}
	
	\title{\texttt{correlcalc}:  A `Generic' Recipe for Calculation of Two-point Correlation Function}
	
	\author{Yeluripati Rohin\fnref{myfootnote}}
	\address{Department of Physics \& Astrophysics, University of Delhi, Delhi}
	\ead{yrohinkumar@gmail.com}
	
	
	
	\begin{abstract}
	 This article provides a method for quick computation of galaxy two-point correlation function(2pCF) from redshift surveys using \texttt{python}. One of the salient features of this approach is that it can be used for calculating galaxy clustering for any arbitrary geometry (or Cosmology) model. Being efficient enough to run fast on a low-spec desktop computer, this `recipe' can be used for quick validation of alternative models and for pedagogical purposes. 	
	\end{abstract}
	
	\begin{keyword}
		Two-point correlation function \sep \texttt{BallTree}\sep Galaxy Clustering \sep redshift surveys \sep 2pCF for alternative models
	\end{keyword}
	
\end{frontmatter}


%
%
\section{Introduction}
The two-point correlation function (2pCF from here on) is one of the vital statistics that can be obtained from the galaxy redshift surveys. 2pCF provides information about the galaxy clustering in redshift space and in turn provides crucial observables such as Baryon Acoustic Oscillation peak position (BAO peak)\cite{eisenstein2005detection}, structure growth rate\cite{peebles1980large} and test of geometry of the Universe through Alcock-Paczynski test\cite{li2016cosmological}.


The brute-force calculation of 2pCF is \textit{O}($N^2$) process\cite{ponce2012application} as one needs to calculate pair-wise distances of data points and bin them. They get more computationally intensive as histogram construction time gets added to the time taken to compute distances as per the assumed Cosmology model\cite{cardenas2013concurrent}. This demands code optimization for multiple-CPUs, clusters and GPUs to decrease the computational time\cite{ponce2012application}\cite{bard2013cosmological}\cite{cardenas2013calculation}. A. Moore \textit{et. al.} \cite{moore2001fast} was one of the first to propose application of nearest neighborhood methods (specifically KDTree) in Cosmology for `fast' calculation of N-point statistics. All the modern codes for 2pCF use nearest neighbor methods in combination with CPU/GPU computational optimizations.

 Many tools and software packages based on these faster implementations to calculate  2pCF exist\cite{reid2013cosmoxi2d}\cite{hearin2016high}\cite{cutebox}\cite{sinha2017corrfunc} and the need for a new software may be questioned. Two primary reasons for this can be explained as follows. First, all of the implementations using nearest neighborhood methods limit themselves to Euclidean distance metric\cite{hearin2016high}\cite{cutebox}\cite{sinha2017corrfunc}\cite{astroML}. While these can be used for a multitude of flat cosmologies, those studying alternative models dealing with non-Euclidean spaces have no generic codes available to obtain correlation statistics. As a result, they resort to using the model-independent methods/parameters to constrain their models\cite{lopez2014alcock}\cite{melia2017alcock}. Second, most of the existing codes are designed with box geometry in mind meant for evaluating the N-body simulations and not for complicated survey geometries. A beginner looking to extract 2pCF from redshift surveys like SDSS/BOSS, has no easy way to start as it needs multiple steps ranging from random catalog creation for survey geometry, defining distance metrics to final measurement of 2pCF.

 \texttt{correlcalc} aims to fill this gap with an easily accessible generic python code for a beginner with step-by-step explanation of the procedure to extract 2pCF statistics from a large scale catalog. The goal of this work is two-fold. First, to present a generic `recipe' to compute 2pCF for any arbitrary cosmology model of any geometry. Second, to compute galaxy 2pCF quickly and reliably with minimal computational resources (on a typical laptop/desktop).

Organization of the content in this paper is as follows. After listing a few basic formulae needed for computation of distances (\ref{cdist}) \textit{etc.} Section \ref{theory} introduces nearest neighborhood methods and their popular implementations in \texttt{python}(\ref{nnmethods}). The section on algorithm(\ref{algo}) sequentially covers the important aspects of the recipe \textit{viz.} data preparation(\ref{datprep}), random catalog creation(\ref{randprep}), data visualization(\ref{datvisu}), defining custom distance metrics(\ref{custmetric}), computing data pairs into bins to finally calculate 2pCF(\ref{ccnn}) and anisotropic 2pCF(\ref{3d2pcf}). After the discussion of algorithm, the article concludes with results(\ref{results}) obtained using this recipe for DR7 VAGC\cite{kazin2010baryonic} catalog with brief discussions. \texttt{python} implementation of the algorithm is available in the form of python package hosted on github\cite{correlcalc} (also on ASCL and PyPI).

\section{Theory}\label{theory}
 The two-point correlation function ($\xi$) is defined\cite{peebles1980large} by the joint probability of finding an object in both of the volume elements $\delta V_1$ and $\delta V_2$ at separation $\chi_{12}$. It measures the excess probability of finding a couple of galaxies separated by spatial distance $\chi_{12}$ (or angular distance $\theta$) in comparison to the galaxies separated by the same distance or angle in a Poisson random distribution.

\begin{equation}
\delta P = n^2 \delta V_1 \delta V_2 [1+\xi(\chi_{12})]
\end{equation}

Here $n$ is number density of the objects.

\subsection{Calculation of 2pCF}

To empirically calculate 2pCF we can use any of the common estimators listed in Vargas \textit{et. al.} \cite{vargas2013optimized} such as the generalized Landy-Szalay estimator \cite{landy1993bias} given below in equation \ref{lsest}. 


\begin{equation}
\xi (s ;z)=\frac{DD-2DR+RR}{RR}\label{lsest}
\end{equation}
	
Here $DD$, $RR$, $DR$  are the normalized values of number of data--data ($dd(s)$), random--random ($rr(s)$) and data--random($dr(s)$) pairs in the specific co-moving distance radius. 

\begin{equation}
DD = \frac{dd(s)}{N_d(N_d-1)/2}~~~ RR =  \frac{rr(s)}{N_r(N_r-1)/2}~~~ DR = \frac{dr(s)}{N_d.N_r}
\end{equation}

Here, $N_d$ and $N_r$ are no. of galaxies (data points) and no. of points in the random catalog respectively. 



\subsection{Calculation of distance between two galaxies}\label{cdist}
Galaxy surveys cannot directly measure the distances of individual galaxies. Instead, they measure redshifts of the objects. Assuming isotropic and homogeneous expansion of the Universe and a cosmology model a priori (fiducial model), we convert these redshifts($z$) to co-moving distances ($\chi(z)$) using the following equation for standard $\Lambda CDM$ model. For other models like $wCDM$, co-moving distance may be similarly defined.

\begin{equation}
\chi (z) = \frac{c}{H_0}\int_0^z\frac{dz'}{E(z')}
\end{equation}

where
\begin{equation}
\label{eq:ez}
E(z)\equiv\sqrt{\Omega_{\rm M}\,(1+z)^3+\Omega_k\,(1+z)^2+\Omega_{\Lambda}}
\end{equation}

Distance between any two points with given redshift ($z$) and angular position ($RA$ and $DEC$) can be found using simple triangle law for flat geometry. For an arbitrary geometry, distance between $P_1(z_1,RA_1,DEC_1)$ and $P_2(z_2,RA_2,DEC_2)$) can be written as\cite{matsubara2004correlation}
\begin{eqnarray}
{\chi_{12}} =
\left[
{S}^2(z_1) + {S}^2(z_2) -
2 C(z_1) C(z_2) S(z_1) S(z_2) \cos\theta - 
\Omega_k {S}^2(z_1) {S}^2(z_2) \left(1 + \cos^2\theta\right)
\right]^{1/2}
\label{dist}
\end{eqnarray}
Where $\chi(z)$ is co-moving distance at $z$ and $\Omega_k$ is curvature defined as $\Omega_k=\Omega_M+\Omega_\Lambda-1$
\begin{eqnarray}
S(z) &\equiv&
S_K[\chi(z)]
\nonumber\\
&=&
\left\{
\begin{array}{ll}
(-\Omega_k)^{-1/2} {\rm sinh}\left[(-\Omega_k)^{1/2} \chi(z)\right], &
(\Omega_k < 0), \\
\chi(z), & (\Omega_k = 0), \\
\Omega_k^{-1/2} \sin\left[\Omega_k^{1/2} \chi(z)\right], &
(\Omega_k > 0),
\end{array}
\right.
\label{eq2-12a}\\
C(z) &\equiv&
\frac{dS_K}{d\chi} (z) =
\left\{
\begin{array}{ll}
\cosh\left[(-\Omega_k)^{1/2} \chi(z)\right], &
(\Omega_k < 0), \\
1, & (\Omega_k = 0), \\
\cos\left[\Omega_k^{1/2} \chi(z)\right], &
(\Omega_k > 0).
\end{array}
\right.
\label{eq2-12b}
\end{eqnarray}
\begin{figure}[h]
	\begin{center}
		\includegraphics[scale=0.6]{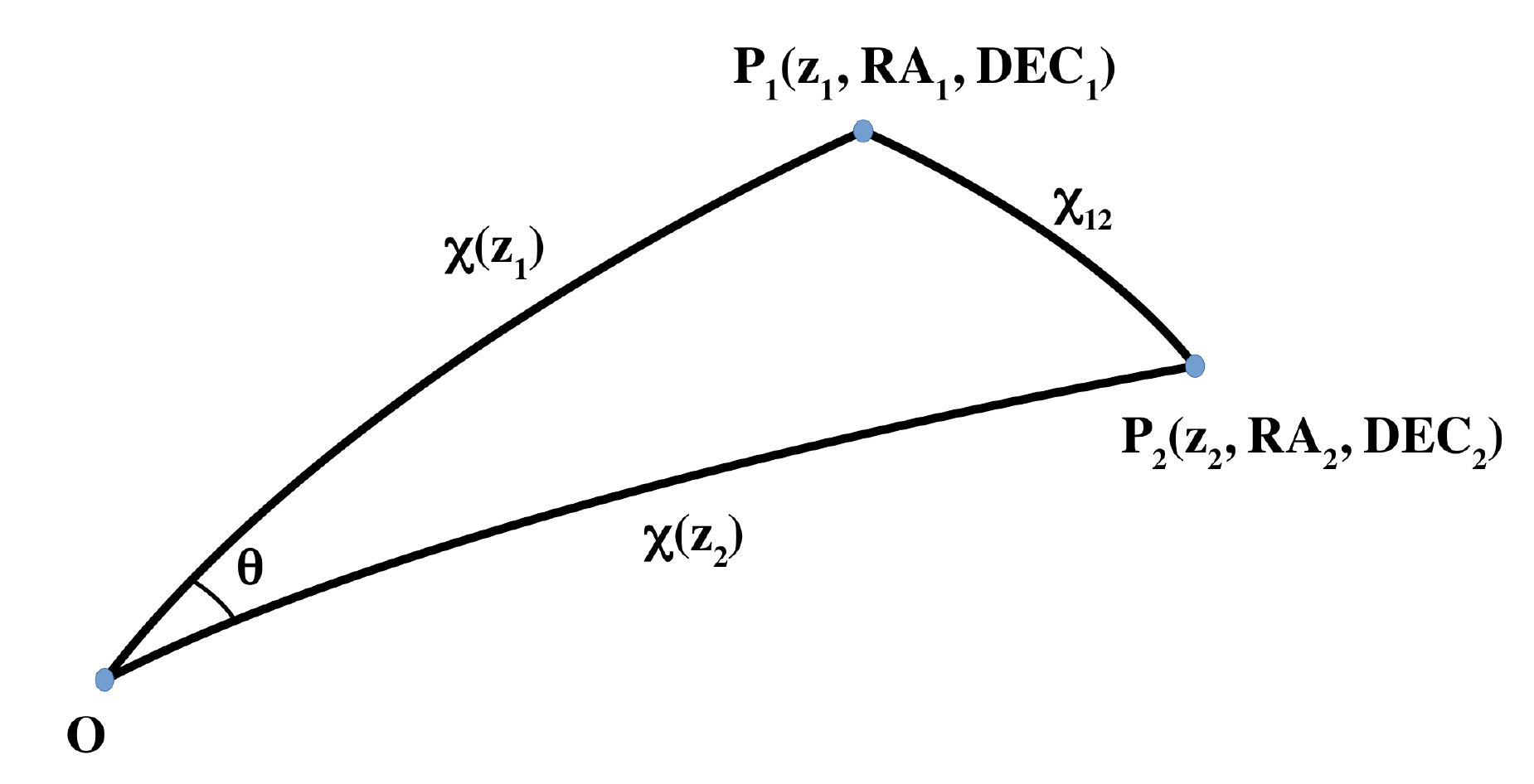}
		\label{fig1}
		\caption{Geometry of the relative position of the observer and the two points. The space is	not necessarily Euclidean.}
	\end{center}
\end{figure}

\subsection{Calculation of LOS distance}\label{pdist}

The line-of-sight distance between two galaxies separated by $\Delta z =z_1-z_2$ is dependent on the Hubble parameter at the average redshift ($z$) as 

$$s_\parallel = \frac{c\Delta z}{H(z)}$$

\begin{figure}[ht]
	\begin{center}
		\includegraphics[width=1.0\columnwidth]{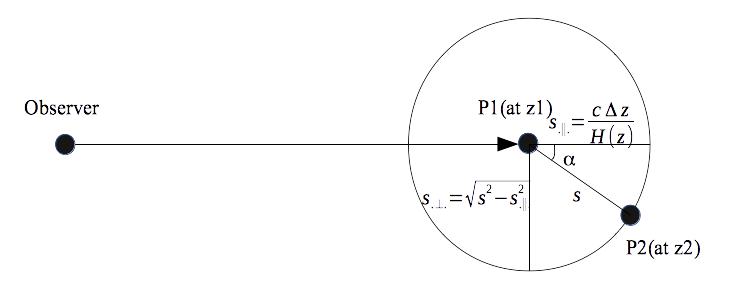}
		\label{parper}
		\caption{Distance measures for 3D 2pCF}
	\end{center}
\end{figure}

Perpendicular to the line-of-sight distance can be calculated using Pythagoras theorem assuming locally flat geometry or using angular diameter distance ($D_A$). We use the former and write ($s_\perp$) as

$$s_\perp = \sqrt{s^2-s^2_\parallel}$$ where $s$ is the total distance between the two galaxies calculated using eqn. \ref{dist}. 

Angle between the two galaxies from the LOS axis ($\alpha$) can be written as 

$$\alpha = \cos^{-1}(\mu)$$ where $\mu = s_\parallel/s$. Two-point correlation as a function of distance($s$) and $\mu$ ($\xi(s,\mu)$) is very useful in calculation of redshift distortions ($\beta$ parameter) as it can be expanded as multipoles of $\mu$.

\subsection{Nearest Neighbor Methods}\label{nnmethods}

Formally, the nearest-neighbor (NN) search problem is defined\cite{knuthart} as follows: given a set $S$ of points in a space $M$ and a query point $q \in M$, find the closest point in $S$ to $q$. 

Nearest neighbor search (NNS), is the optimization problem of finding the point in a given set that is closest to a given point. Closeness is typically expressed in terms of a dissimilarity function: the less similar the objects, the larger the function values. Most commonly, $M$ is a metric space and dissimilarity is expressed as a distance metric. Between two points, a distance metric must be symmetric and should satisfy the triangle inequality. Any arbitrary metric one likes to define must follow these two conditions. Often, $M$ is taken to be the $d$-dimensional vector space where dissimilarity is measured using the Euclidean distance, Manhattan distance or other distance metric\cite{scikit-learn}. 

Brute-force method is to simply calculate distances between all the pairs and find the closest points by binning them based on the `closeness'. This process is inefficient for large no. of data points as the computational time scales as $N^2$ where $N$ is the no. of data points. One can, however, deploy some machine learning to analyze the data/node structure in terms of a given metric and efficiently find nearest neighbors based on the `learning'. 

Several space-partitioning methods have been developed for solving the NNS problem. The simplest example is the KDTree, which iteratively bisects the search space into two regions containing half of the points of the parent region. Queries are performed via traversal of the tree from the root to a leaf by evaluating the query point at each split. Depending on the distance specified in the query, neighboring branches that might contain hits may also need to be evaluated. For constant dimension query time, average complexity is \textit{O}($\log N$)\cite{jakedvp}. 

\begin{figure}[h]
	\begin{center}
		\includegraphics[scale=0.9]{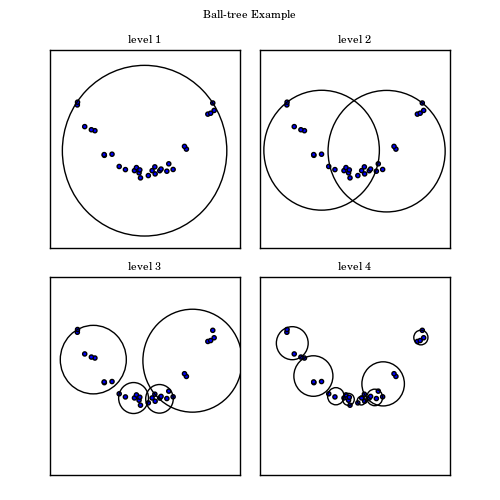}
		\label{balltreeeg}
		\caption{\texttt{BallTree} creation steps (figure taken from \cite{astroML})}
	\end{center}
\end{figure}

Most codes that calculate 2pCF (for e.g. CUTE\cite{cutebox}) use KDTree to find the nearest neighbors which leads to very fast results on even ordinary  computers. However, KDTree implementations inherently work in Euclidean space. As our target is to find a generic recipe irrespective of the geometry, we use \texttt{BallTree} for application in Non-Euclidean spaces. BallTree algorithm is similar to KDTree but instead of calculating the median values and assigning space-zones iteratively, BallTree divides space into leaves with varying radius. Example in figure \ref{balltreeeg} depicts how a simple Ball tree partition of a two-dimensional parameter space is created step-by-step.



Once an indexing structure is created, one can find the number of points near any random node (within a specified distance) by simply counting the leaves of the tree. Based on the `leaf-size' of the tree one might need to calculate individual distances by brute-force method near the edges of the tree. More details on the implementation of these algorithms can be found at scikit-learn (sklearn) documentation.






\section{Algorithm}\label{algo}
\begin{figure}[h]
	\begin{center}
		\includegraphics[scale=0.2]{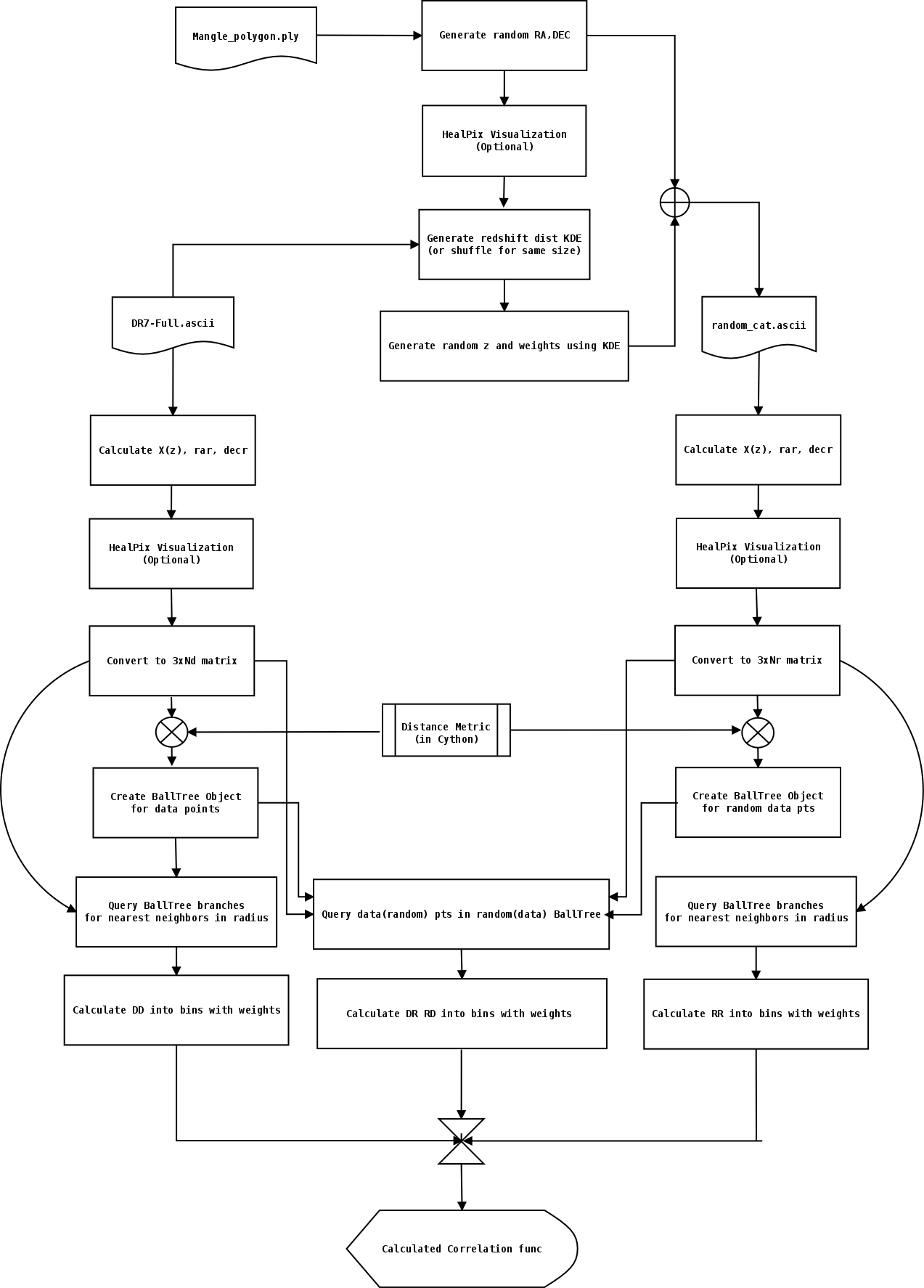}
		\label{fig2}
		\caption{Flow chart of the `generic' 2pCF computation recipe}
	\end{center}
\end{figure}

First of all, we take galaxy/quasar data from a redshift survey. In this paper, we take SDSS DR7 VAGC from Kazin \textit{et. al.}\cite{kazin2010baryonic} to demonstrate the procedure. Typical data from a redshift survey contains list of galaxies with their observed redshift (z), Right ascension (RA) and declination angles (DEC) that provide the position of each galaxy. Often, a value added catalog limits the magnitude and chooses a specific type of galaxy such as Luminous Red Galaxies - LRGs in DR72 VAGC. There are more parameters in a typical catalog such as magnitude, survey completeness etc. In this paper, we shall only make use of the redshift, angular positions and radial weights of each galaxy from the catalog. Survey geometry and angular distribution are taken in mangle polygon file formats (\texttt{.ply}) provided by the respective survey data release. Value added catalogs such as the ones taken from (non-official) SDSS DR72 \cite{dr72vagc} also provide random catalogs. Thye are typically much larger than the galaxy catalog (15 -- 30 $\times$). Hence, the computation time depends mainly on the size of the random catalog. The procedure provided in this paper can certainly be run on these random catalogs. However, as we will see later (\ref{randprep}), to reduce the computational effort we can create a smaller random catalog without much loss of accuracy in the final 2pCF result. Figure \ref{fig2} depicts flowchart of the recipe. Implementation of the algorithm with block-by-block explanation and code is available in the form of a \texttt{python} package \texttt{correlcalc} on github\cite{correlcalc}. 

\subsection{Data Preparation}\label{datprep}
After installing relevant packages, we need to obtain data from a galaxy catalog. In this work, we are demonstrating with the DR72 VAGC taken from Kazin \textit{et. al.} \cite{kazin2010baryonic}. One can also create a large scale structure galaxy catalog by following instructions given in the SDSS LSS catalog creation tutorial\cite{sdsslsscat}. After obtaining the catalog with redshift($z$) and angular co-ordinates, we calculate the co-moving distance as per the model of our choice (fiducial cosmology). Next the data is loaded into the RAM as $3\times N_d$ matrix $[s,rar,decr]$ where $s$ is co-moving distance, $rar$ and $decr$ are right ascension and declination in radians. While these calculations can be done as part of the distance metric to be defined in section \ref{custmetric}, it is faster to calculate them first and keep distance calculation minimal. It is also wise to save the data in $3\times N_d$ matrix format in an ascii file or in pickled data form for easy access.
%

\subsection{Random Catalog Preparation}\label{randprep}
If we use the random catalog given by the survey, we follow the same steps to load random catalog into RAM in form of $3\times N_r$ matrix with $[s,rar,decr]$ values. However, it is computationally intensive to calculate correlation in Landy-Szalay method (and other methods which need random-random correlation) for high no. of random points. So, we can create a smaller random catalog which can give reasonably accurate results in comparison to the standard random catalog. (Only exception is probably Davis \& Peebles method\cite{vargas2013optimized} that only needs DD and DR, in which case this recipe will yield even quicker results.)

All the major surveys provide the survey geometry in form of angular masks in \texttt{fits} and/or \texttt{ply} formats. We use mangle polygons \cite{hamilton2004scheme} \cite{swanson2008methods} given by the galaxy surveys to create random points that fall within the survey geometry.

We use \texttt{pymangle} -- a python wrapper to a faster \texttt{C/C++} code to manipulate mangle angular masks\cite{pymangle}. It can create a random catalog with a specified no. of data points that follow the angular masks provided by the survey. Mangle gives random points in the angular domain only providing RA and DEC. We will also need redshift values to be assigned to these random points. These redshifts need to follow the survey radial distribution to create a matching random catalog. If we are creating a random catalog that is of the same size as the data catalog then it is easier to re-use the redshifts of the data and shuffle them to assign to the random catalog points. This ensures the radial distribution of the random points to be the same as galaxy catalog. However this can result in reduced accuracy due to suppressed radial modes. In stead it is better to create a catalog that has same distribution as the input data catalog\cite{kazin2010baryonic}.

If we are to create a random catalog of different size (recommended to use at least $\geq2\times$ of the data catalog to reduce shot noise.) we can plot the histogram or Kernel density estimator(KDE) of the data redshift distribution and create a random catalog that mimics this distribution. As KDE is a better choice than using a histogram, we create a random distribution points for redshifts using the same. \texttt{pymangle} though faster, is not parallelized to generate random angular points quick enough for high number of data points. (It can take more than 2 hours to create a random catalog of $2\times$ size for DR7, this process is a bottle-neck). Once the catalog is created and stored in an ascii file it can be reused for different realizations and mock catalogs. To minimize the variance of clustering measurements for an inhomogeneous sample, we assign radial weights to each random point created. While calculating the pair counts, we assign to each data point a radial weight of $1/(1 + n(z)·P_w)$\cite{feldman1993power}, where $n(z)$ is the radial selection function and $P_w \sim 10^4h^{-3}Mpc^3$\cite{kazin2010baryonic}



\subsection{Healpix Visualization}\label{datvisu}

\begin{figure}[h]
	\includegraphics[scale=0.6]{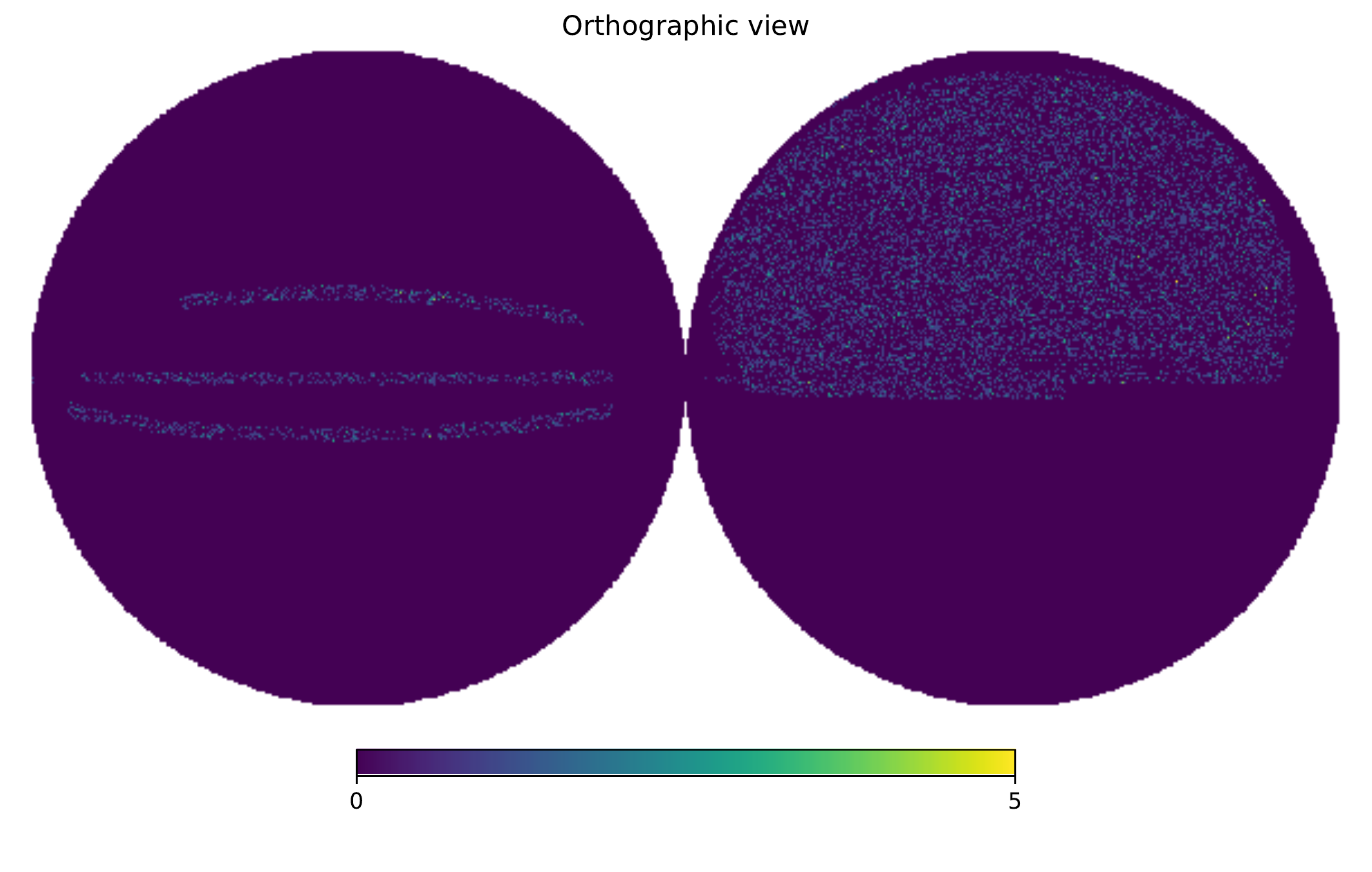}
	\caption{Generated random catalog with DR72 mask - \texttt{orthview}}
	\label{fig3}
\end{figure}

It is often useful to have a visual representation of the galaxy catalog distribution. HealPix\cite{2005ApJ...622..759G} (and the wrapper healpy) are extremely useful to visualize the galaxy surveys. After creating the random catalog it is good to cross-check if they all indeed follow the supplied geometry. Using Healpix, we plot the random catalog containing RA and DEC data with points created as per the mask as shown in figure \ref{fig3}. 

Depending on the NSIDE parameter, each pixel covers some angular area in which galaxies of relevant RA, DECs fall. Pixels are assigned values of total weights added of all the galaxies. Iterating this process to calculate pixel data for all the galaxies (or random points) in the survey area creates a pixelized map of the point distribution.

\begin{figure}[h]
	\includegraphics[scale=0.6]{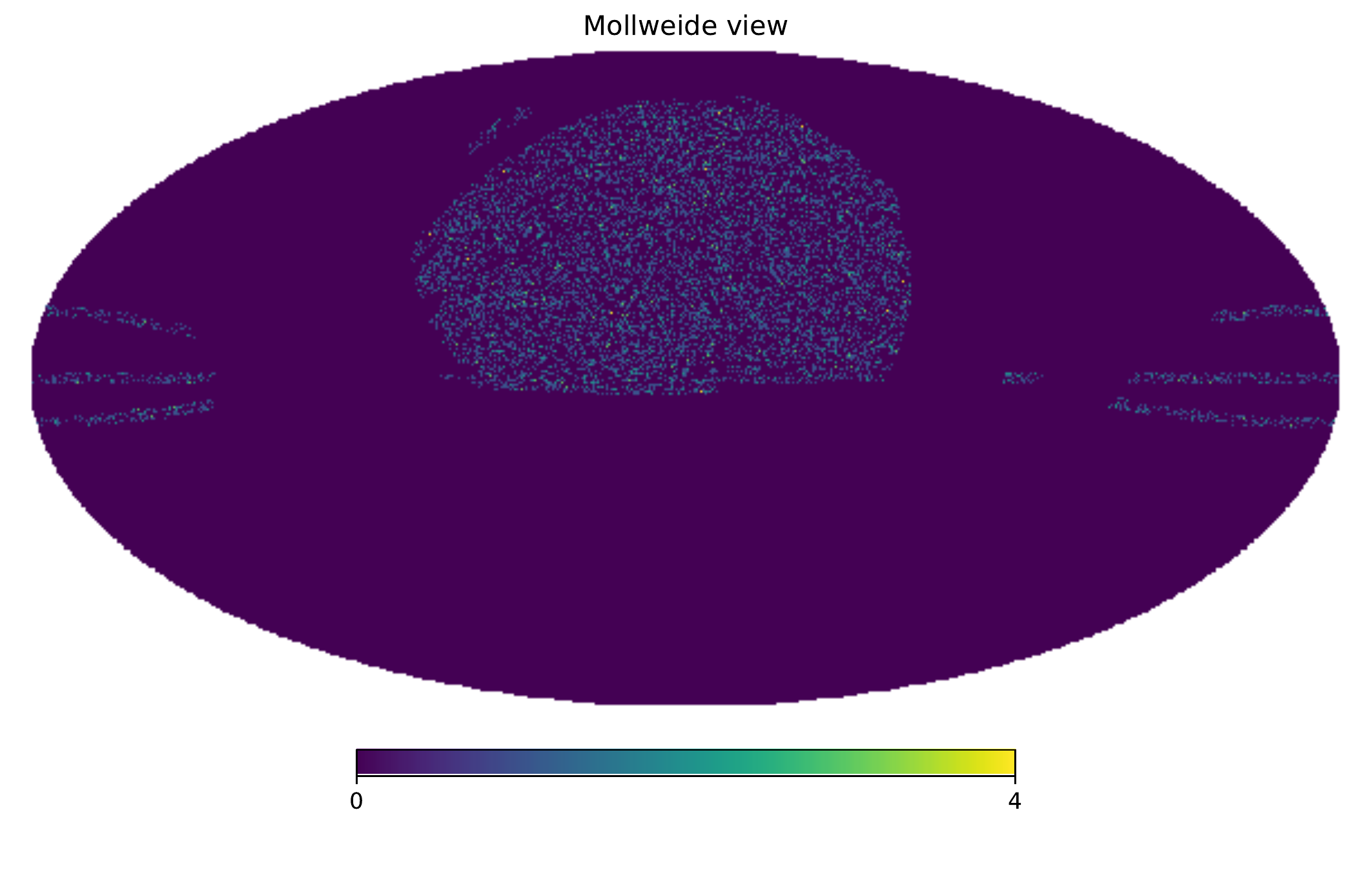}
	\caption{Generated random catalog with DR72 mask - \texttt{mollview}}
	\label{randmoll}
\end{figure}

\subsection{Defining distance metric}\label{custmetric}
Implementations of nearest neighbor methods in popular packages like \texttt{sklearn} \cite{scikit-learn} offer many distance metrics. Among them, only Euclidean metric is useful for application in the study of Large Scale Structures. Among the nearest neighbor implementations, only \texttt{BallTree} implementation of scikit-learn contains a provision to define `custom' metric in stead of the default metrics. This metric should not be confused with FRW metric of cosmology. Here metric definition is for calculation of distances between two points given their co-ordinates (in this case co-moving distance and angular positions).

As described in section \ref{theory}, we use the formulae given in Matsubara\cite{matsubara2004correlation} to define custom metrics for arbitrary geometry. It is worth noting that calculation of co-moving distances done in section \ref{datprep} is already dependent on the cosmology of our choice. As math functions are inherently faster in \texttt{C} than in \texttt{python}, we use \texttt{Cython}(a \texttt{C} implementation of \texttt{python}) to improve the speed of computation. We know that angle between two points on a sphere (points need not be at the same distance since we are only calculating angle between them. It can be projected onto a surface - celestial sphere) is 
\begin{equation}
\cos\theta=\sin(DEC1)\sin(DEC2)+\cos(DEC1)\cos(DEC2)\cos(RA1-RA2)
\end{equation}

It is advisable to use square of the distance instead of the square root to reduce the additional computational effort that is needed to calculate \texttt{sqrt} function. In the later steps, binning can be done based on distance squares. For closed and open Universe cases, distance units become important as the values inside $\sin$ and $\sinh$ need to be dimensionless. We use $c/H_0$ as the default units for distance measurement where $c$ is velocity of light and $H_0$ is Hubble parameter at the present epoch. Also, one needs to take into account that the formulae given in section \ref{cdist}(from \cite{matsubara2004correlation}) are in natural units ($c=1$). So in order to get dimensionless numbers inside $\sin$ or $\sinh$ we need to multiply the co-moving distance with an appropriate factor assuming the current Hubble expansion rate. 
After compiling the \texttt{Cython} code into a \texttt{python} module, we can import the same in the main code and use it as a custom metric in the BallTree creation.

\subsection{Creating \texttt{BallTrees} and calculating 2pCF}\label{ccnn}

We now create a BallTree using the custom metric for data catalog and random catalog separately. In creating the BallTree we can find the optimal leaf size to improve computation time/efficiency\cite{jakedvp}. Optimizing \texttt{leaf\_size} parameter gives best trade-off between node query time and brute-force distance computation. For more details on benchmarking and optimization parameters nearest neighbor algorithms check the article in reference \cite{jakedvp}. 

After creating a BallTree object, we can use \texttt{query\_radius} method to find the number of data points within a given radius. There is also a faster method-\texttt{two\_point\_correlation} (though not parallel) which calculates auto correlation function of the data within given bins. We use this for calculation of $DD$,$RR$ and $DR$ ($RD$) by subtracting the no. of pairs in the smaller bins progressively. We do this using \texttt{numpy.diff} method. To calculate data-random correlation ($DR$), we need to find the random points in the data-BallTree or vice-versa. This gives us $DR$. As the nearest neighbor methods are approximate methods, while counting the pairs in a given radius some of the nearest neighbors may not be counted while few others may get included. This can result in slight deviation (especially at large scales) in $DR$ and $RD$ values as they use different trees for calculation of cross-correlation. Hence, we normalize this effect by taking an average of $DR$ and $RD$ to use it in the 2pCF formula.


The above mentioned approach assigns a pair to the relevant bin by adding ``1" every time a pair is found within the specified distance. However, in surveys such as SDSS we often need to add relevant weights instead of adding 1's. These weights have to be systematically added to the bins to obtain accurate 2pCF. For more details on different types of weights used in the preparation of LSS catalog please refer to the ``Weights" section of SDSS tutorial \cite{sdsslsscat}.

To achieve this weighted sum, we can either patch the existing  \texttt{two\_point\_correlation} method in \texttt{sklearn} or define a method that can do it with similar efficiency. Here, we follow the latter approach and create methods to find autocorrelation and cross-correlation of a distribution with previously defined metrics. Just as in the case of brute-force approach, one can calculate all the pairwise distances to bin them to obtain $DD$, $RR$ and $DR$. But instead of calculating distances for all the pairs, we narrow down the search of pairs to the maximum distance needed by the provided bins. For e.g. if we like to calculate 2pCF up to 200Mpc distance, we first need to find the pairs that are in 200Mpc radius and calculate distances only for the pairs that fall in this radius. We repeat this process for each data/random point removing them one by one after the neighbor distances are calculated. This ensures we do not count the same pairs multiple times. 

Implementing this strategy is straightforward once a \texttt{BallTree} is created for data and random points. To calculate the autocorrelation ($DD$ and $RR$), we first create a \texttt{BallTree} of the data and iterate through all the points in that dataset to find the nearest neighbors within the given radius (max. value of the bins provided). This narrows down the no. of pairs to a smaller manageable number. We can then calculate distances between the chosen point and its neighbors with the specified custom metric using \texttt{scipy.spatial.distance.cdist} and bin the distances as per the weights of those neighbors (using \texttt{numpy.histogram}'s weights option). This process when iterated over all the points gives us autocorrelation. This iterative loop can be parallelized to achieve speed of computation.


To calculate cross-correlations ($RD$ for e.g.) we iterate over random catalog points in the data \texttt{BallTree} to find nearest neighbors within the radius. We then calculate distances between the nearest neighbors of random catalog points in the data point catalog to bin them with (data) weights to calculate the cross-correlation. 
 
Having obtained $DD$, $RR$ and $DR$, we can find any of the common estimators as described in the theory section. For quick computation using Landy-Szalay estimator on low-spec machines, we can create a smaller (typically 2$\times$) size random catalog\ref{randprep}. For quick results, we can also use Davis - Peebles estimator as it only needs $DD$ and $DR$. It is easy to compute as it avoids calculation of $RR$ for random catalogs of large size. Although it is not a very accurate estimator\cite{vargas2013optimized}  at scales above 100-125Mpc (for say BOSS geometry), it is good enough for pedagogical purposes as it can compute 2pCF very fast and yet show basic features such as power-law of 2pCF at small scales and BAO peak. In situations where we need more accurate results we use the Landy-Szalay estimator \cite{landy1993bias} given in equation \ref{lsest}. 

\subsection{Calculating anisotropic 2pCF}\label{3d2pcf}

Anisotropic(3D) 2pCF is more useful as it can separate evolution effects in line-of-sight and angular diameter distance in the perpendicular direction. To calculate anistropic 2pCF, we need to find distances using two metrics - typically parallel to the line-of-sight(LOS) - $s_{\parallel}$ and perpendicular to the LOS - $s_\perp$. We can also use co-moving distance (s) and cosine of the angle from LOS ($\mu=\cos(\alpha)$). Definition of these parameters is shown in figure \ref{parper}

Alternatively, there is $\Delta z$ and $z\Delta\theta$ parametrization for model-independent calculation of anisotropic 2pCF\cite{lopez2014alcock} which can be extended to all general cases by using suitable metrics. This method can also be used to calculate other realizations for small $\Delta z$. We compute anisotropic 2pCF in all the above realizations to demonstrate the algorithm. 

\begin{figure}[h]
	\begin{center}
		\includegraphics[scale=0.2]{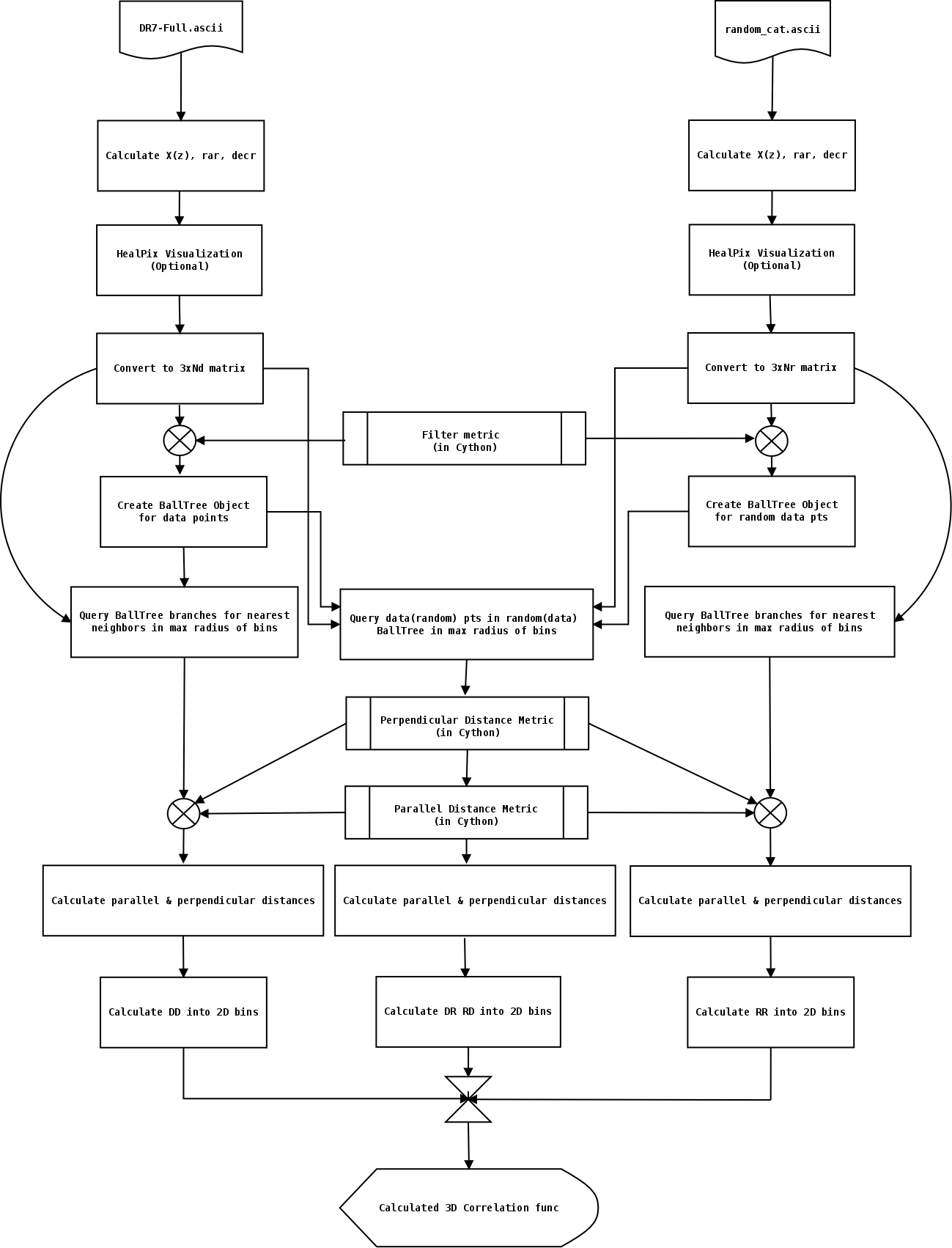}
		\label{3d2pcfalgo}
		\caption{Flow chart of the 3D 2pCF computation recipe}
	\end{center}
\end{figure}

To efficiently calculate anistropic 2pCF, we follow similar approach to the calculation of 2pCF. Since we have two metrics instead of one, we narrow down the the search of pairs using one `filter' metric (typically the radius) and calculate pairwise distances using both the metrics. For e.g. if one metric calculates $\Delta z$ and another $z \Delta\theta$, we first select all the pairs that fall in radius of $\sqrt{(z \Delta\theta)^2 + (\Delta z)^2}$ and then calculate distances using both the metrics. This is similar to the approach taken by Lopez \cite{lopez2014alcock} and it helps in reducing the computational effort. Algorithm to calculate anisotropic 2pCF (also called 3D two-point correlation) is described in figure \ref{3d2pcfalgo}.

\section{Results and Discussions}\label{results}
Following the recipe in section \ref{algo}, we obtain two-point correlations for standard model - flat $\Lambda CDM$ with $\Omega_M=0.3$ and $\Omega_{\Lambda}=0.7$ as fiducial model for SDSS DR7 VAGC catalog and summarized in one plot(\ref{tab:results}). Demonstrating the recipe's capability to work for alternative models, we also obtain 2pCF for open geometry (Milne) and $R_h=ct$ to compare with the standard model (fig.\ref{fig:taba}). To reduce the computational time, a smaller random catalog of $2\times$ size is created. It is remarkable to note that it doesn't cause any significant loss of accuracy in comparison to the default random catalog ($\approx 16\times$) (fig.\ref{fig:tabb}). Hence, this recipe and the associated package can be extremely useful to quickly validate alternative models and can be used for pedagogical purposes. Computation time using $2\times$ random catalog was about an hour to obtain the final results using LS estimator\footnote{Run on i7-2.4GHz quad-core with 16GB RAM}.
We also tested the impact of adding weights instead of doing `+1' in counting pairs. As also confirmed in Kazin \textit{et.al.}\cite{kazin2010baryonic}, there is no significant difference observed in using weights in pair counting as can be seen in fig.\ref{fig:tabc}. For comparison, a plot of with DR3-LRG catalog (original BAO peak detection data) is also presented. We can see the BAO peak clearly around $100 h^{-1} Mpc$ in all the plots.

 We calculated and plotted 3 different realizations of the 3D two-point correlation function. Fig.\ref{fig:tabd} is model-independent plot of 2pCF $\xi(\Delta z, z\Delta\theta)$. $\xi(s_\parallel,s_\perp)$ is plotted as a function of line-of-sight distance ($s_\parallel$) and perpendicular to the LOS ($s_\perp$) in fig.\ref{fig:tabe}. In fig.\ref{fig:tabf}, we plot $\xi(s,\mu)$ as a function of distance($s$) and $\mu$. These match the results obtained by Kazin \textit{et.al.} \cite{kazin2010regarding} as expected.
 
 While the algorithm presented here is usable and extend-able to numerous cosmology models of different geometries, it can be further optimized to achieve better speed. We plan to implement a full Cython code in future revisions of the code. Study of correlation statistics in alternative models using \texttt{correlcalc} will also be explored in the future work(s).
 

{\centering
	\begin{table}[ht]
		\begin{tabular}{cc}
			\begin{subfigure}{0.5\textwidth}\centering\includegraphics[width=1.0\columnwidth]{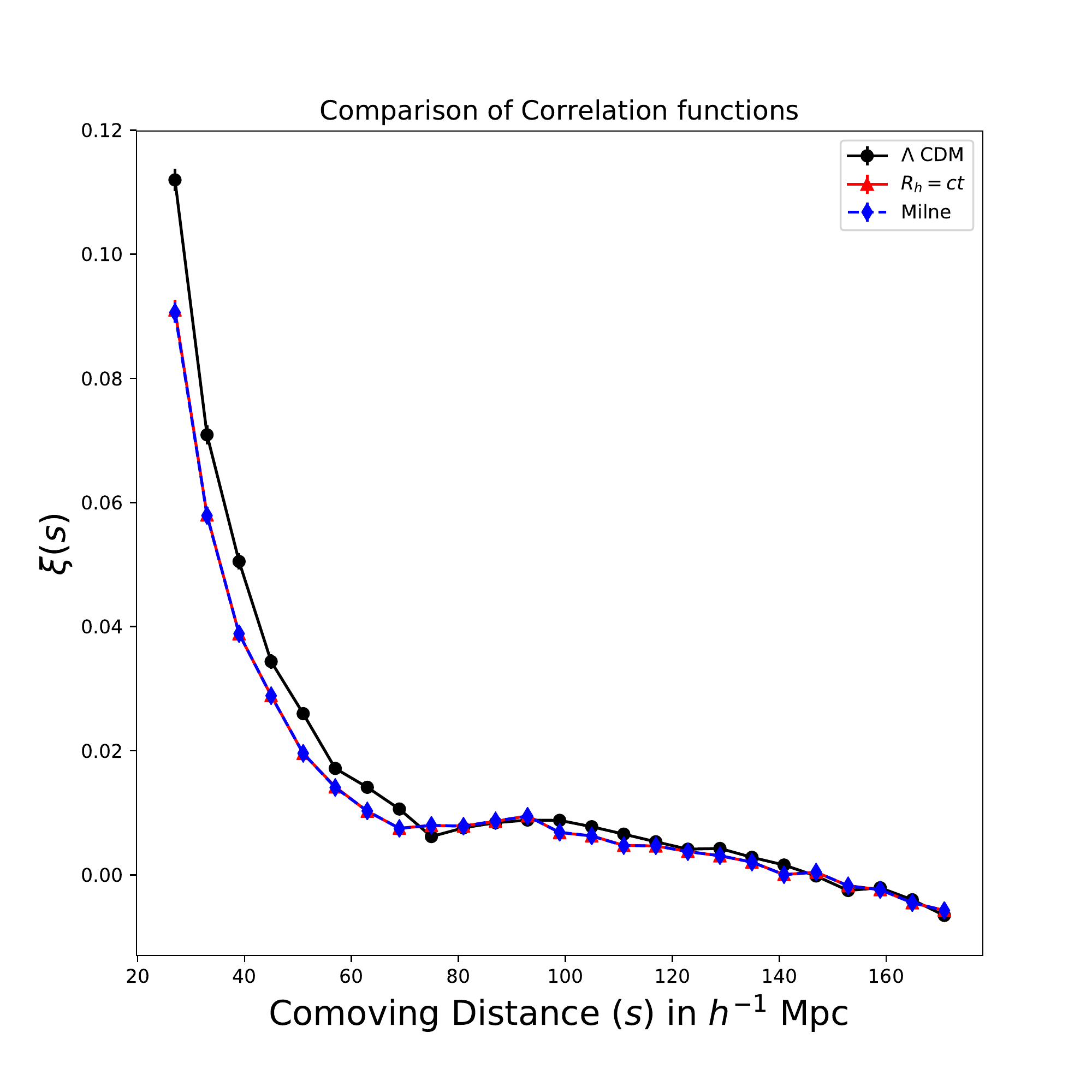}\caption{Figure 1a}\label{fig:taba}\end{subfigure}&
			\begin{subfigure}{0.5\textwidth}\centering\includegraphics[width=1.0\columnwidth]{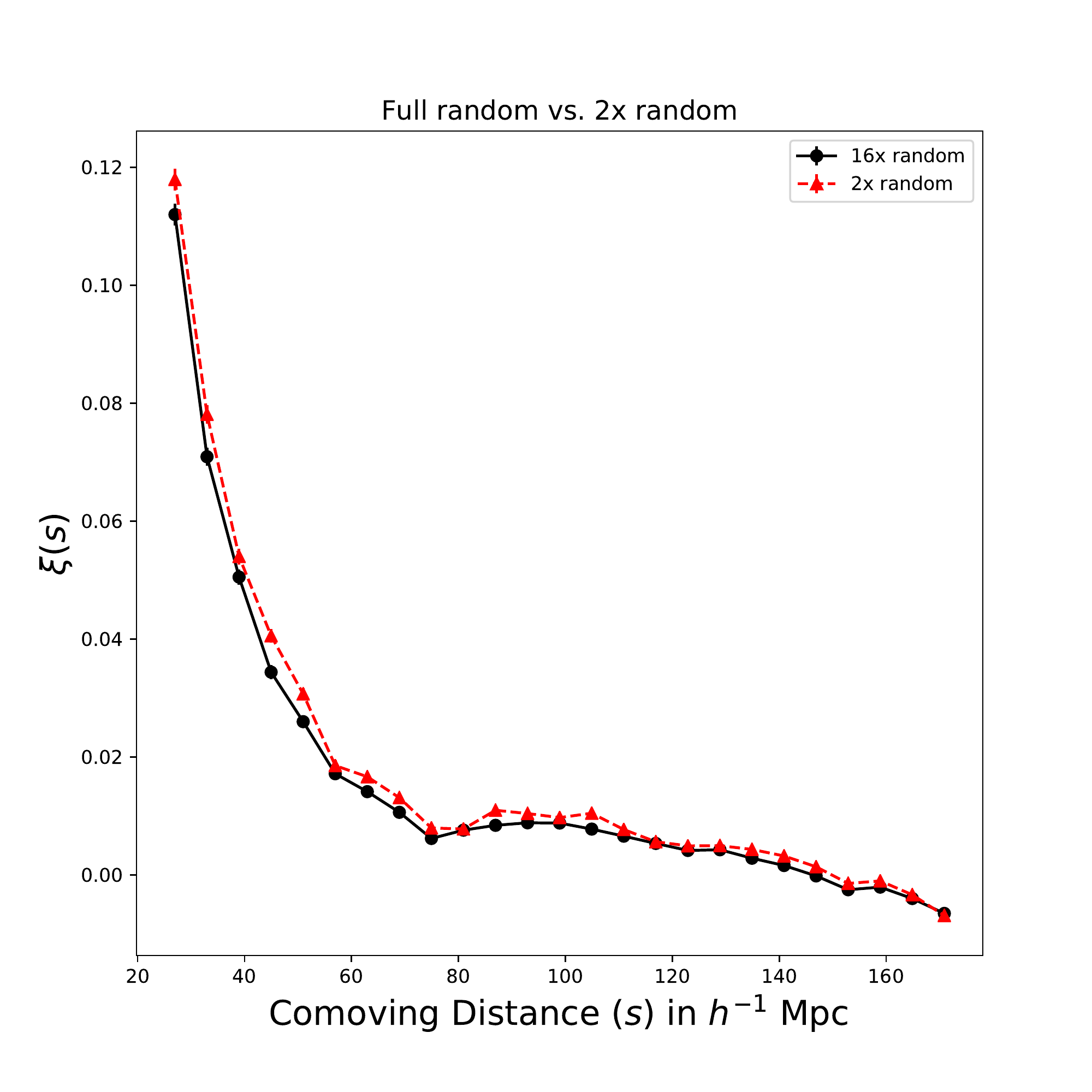}\caption{Figure 1b}\label{fig:tabb}\end{subfigure}\\
			\newline
			\begin{subfigure}{0.5\textwidth}\centering\includegraphics[width=1.0\columnwidth]{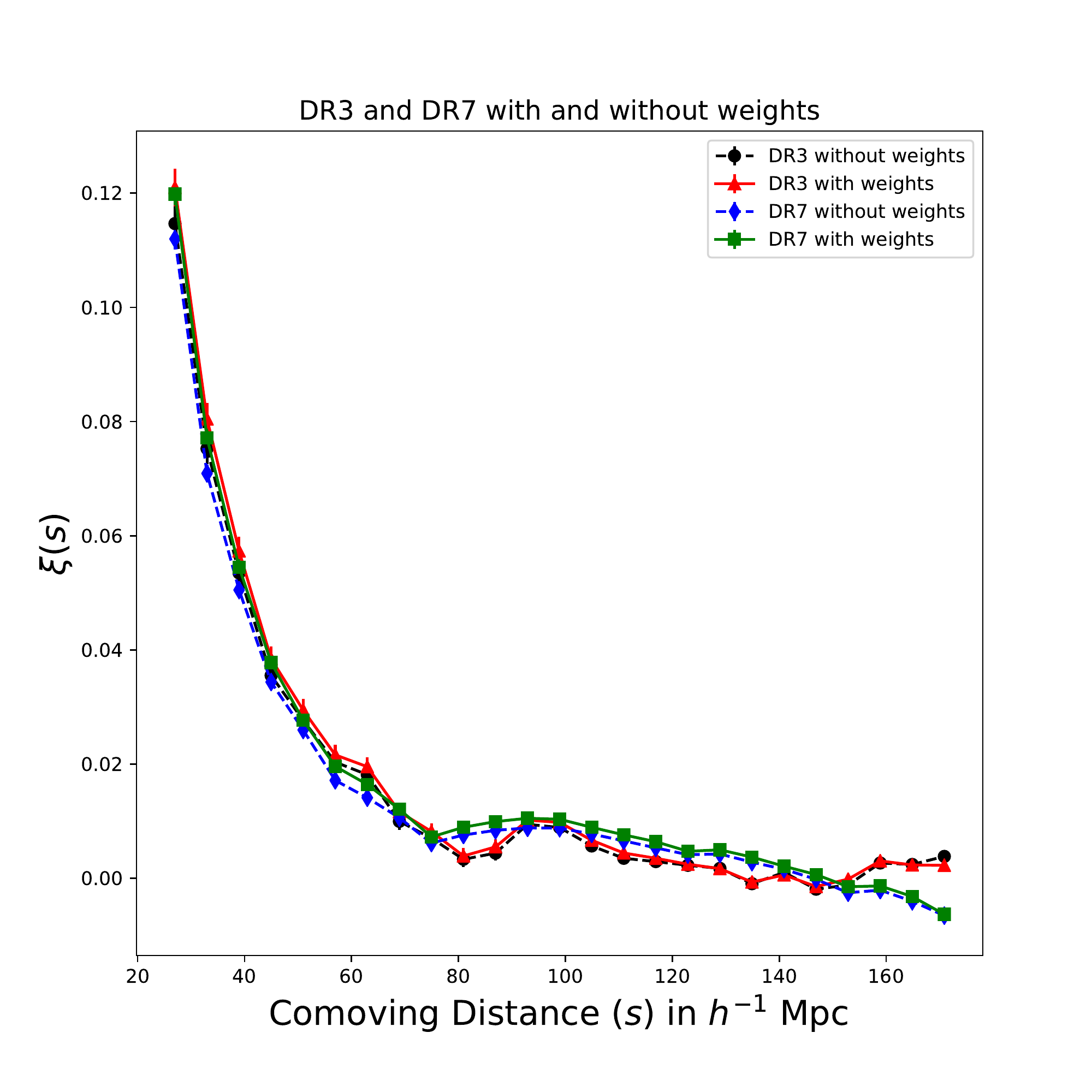}\caption{Figure 1c}\label{fig:tabc}\end{subfigure}&
			\begin{subfigure}{0.5\textwidth}\centering\includegraphics[width=1.0\columnwidth]{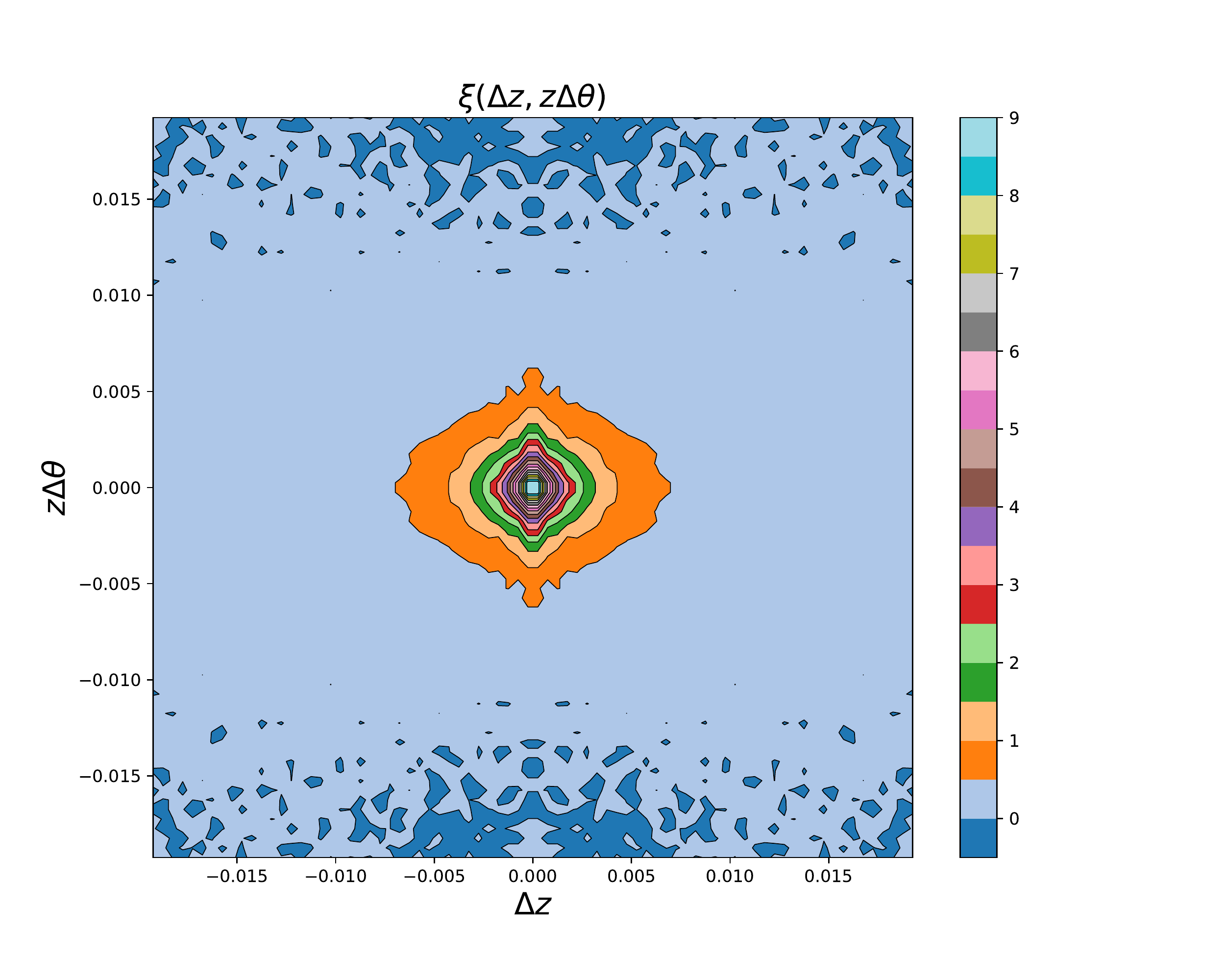}\caption{Figure 1d}\label{fig:tabd}\end{subfigure}\\
			\newline
			\begin{subfigure}{0.5\textwidth}\centering\includegraphics[width=1.0\columnwidth]{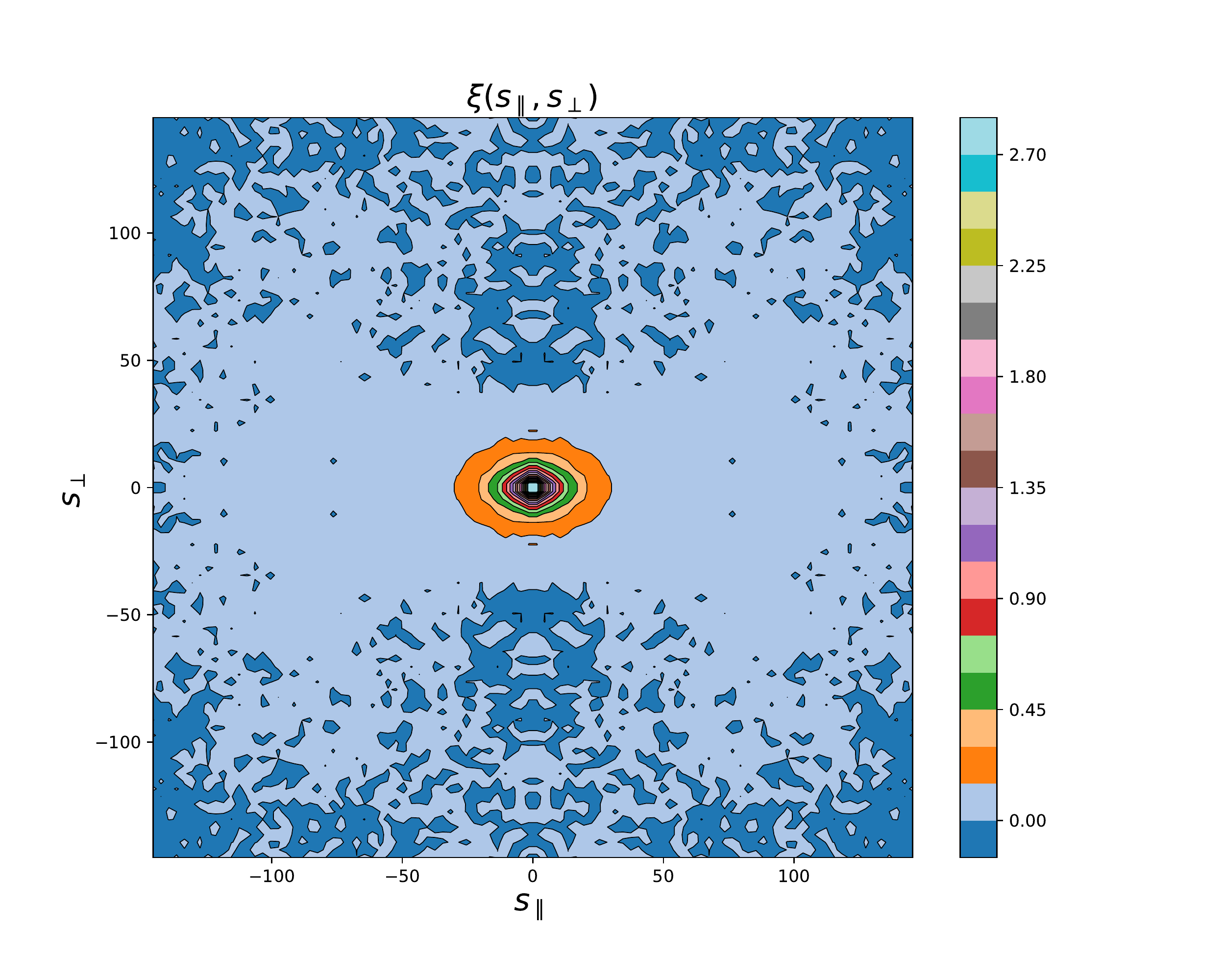}\caption{Figure 1e}\label{fig:tabe}\end{subfigure}&
			\begin{subfigure}{0.5\textwidth}\centering\includegraphics[width=1.0\columnwidth]{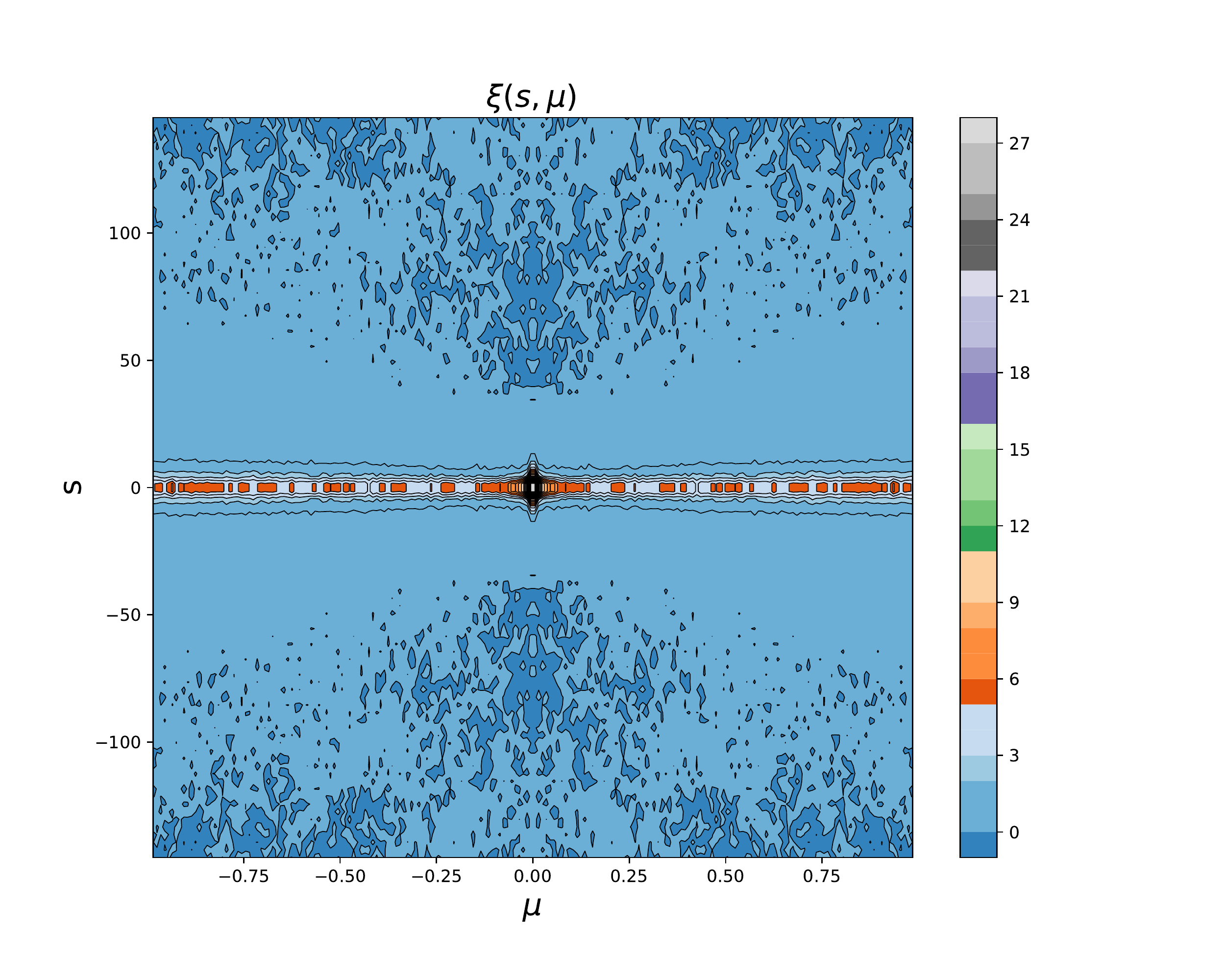}\caption{Figure 1f}\label{fig:tabf}\end{subfigure}\\
		\end{tabular}
		\caption{Two-point correlation functions obtained using \texttt{correlcalc}}
		\label{tab:results}
	\end{table}
}

\section{Acknowledgments}
I thank Prof. Daksh Lohiya \& Prof. Amitabha Mukherjee for their support and encouragement. I am grateful to IUCAA visitor's program dean Prof. K. Subramanian and administration for providing me opportunity to work and avail computational cluster facilities at the institution. Special thanks are due to Dr. Martin Lopez Corredoira, Dr. Aseem Paranjape, Dr. Shadab Alam and Akshay Rana for useful discussions and feedback.
I thank CSIR for providing financial support for my research work through grant no. 09/045(1324) 2014-EMR-I. 
\newpage
\section*{References}
\bibliography{correlcalc}

\end{document}